%Paper: hep-th/9206067
%From: A.A.Tseytlin <A.A.Tseytlin@damtp.cambridge.ac.uk>
%Date: Wed, 17 Jun 92 11:20:14 BST

\input harvmac

\Title{DAMTP - 92 - 36
\ \  (hepth/9206067)
}
{\vbox{\centerline{String cosmology and dilaton}
}}

\centerline{ A.A. Tseytlin
\footnote{$^\dagger$}
{
 On leave of absence from the Department of
Theoretical Physics, P. N. Lebedev Physics Institute, Moscow 117924, Russia.}}
\bigskip
\centerline{\it DAMTP}
\centerline{\it Cambridge University}
\centerline{\it  Cambridge CB3 9EW }
\centerline{{\it United Kingdom}
\footnote{$^*$}{e-mail address: aat11@amtp.cam.ac.uk }}

\baselineskip=20pt plus 2pt minus 2pt
\vskip .3in

%abstract
 We discuss some aspects of string cosmology with an emphasis on the role
played
by the dilaton. A cosmological scenario  based on the assumption that all
spatial dimensions
are periodic so that winding modes play an important role  is reviewed.  A
possibility of  a transition  from a `string phase' to the `standard' cosmology
via  a radiation dominated era and inflation  is  analysed.  We also summarise
some recent results about time dependent solutions of tree level string
equations.

\bigskip \bigskip
\bigskip
\it{To appear in the Int.J.Mod.Phys.A  issue of the  Proceedings of the
International Workshop
 ``String Quantum Gravity and Physics at the Planck Energy Scale", 21-28 June
1992,
Erice, Sicily.  }

\Date{6/92} %replace this line by \draft  for preliminary versions
             %or specify \draftmode at some point

%if you want double-space, use e.g. \baselineskip=20pt plus 2pt minus 2pt

\def \const {{\rm const }}

\def \ha {{1\over 2}}

\def \dvps {{\dot \varphi}^2}
\def \dls {{\dot \lambda }^2}

\def \dop  {\dot {\phi} }

\def \eq#1 {\eqno{(#1)}}
\def \e#1 { e^{#1}}

\def \mi {m_{i}^2}

\def \tmi {{\tilde m}_{i}^2}

\def \dpss {{\vert {\dot \psi} \vert }^2}
\def \pss {{\left| {\psi} \right|}^2}
\def \tpss {{\vert {\tilde \psi} \vert}^2}
\def \dtpss {{\vert {\dot{\tilde \psi}} \vert }^2}

\def \a {\alpha}
\def \b {\beta}

\def \b {\beta}

\def \pa {\partial}

\def \P { \Phi}

\def \fourth { {1\over 4}}
\def \m {\mu}  \def \le { \l_{ \E }  }

\def \ep {\epsilon}
\def \sn {\sqrt N }

\def \dvps {{\dot \varphi}^2}\def \dops {{\dot \phi}^2}
\def \dls {{\dot \lambda }^2}

\def \dop  {\dot {\phi } } 

\def \L {\Lambda}
\def \le {\Lambda}
\def \eq#1 {\eqno{(#1)}}

\def \sf {string frame }
\def \ef {Einstein frame }

\def \g {\gamma}

\def \a {\alpha}

\def \sh { {\sinh}\  }
\def \ch { {\cosh}\  }
\def \th { {\tanh}\  }
\def \cos {\ {\cos} \ }
\def \l {\lambda}
\def \p {\phi}
\def \vp {\varphi}
\def  \g {\gamma}

\def \s {\sigma}
\def \t {\tau}
\def \b {\beta}

\def \pa {\partial}

\def \sqG {\sqrt {-G}}

\def \Gmn {G_{\mu \nu}}
\def \CE {{ U } }
\def \E {{ E }}
\def \CU {{\hat U } }

\def \dl {\dot \lambda}
\def \ddl {\ddot \lambda}
\def \dvp {\dot \varphi}
\def \ddvp {\ddot \varphi}
\def \sm {\sum_{i=1}^{N}}
\def \nineit {\it }
\def\np  { {\nineit  Nucl. Phys. }}
\def \pl { {\nineit  Phys. Lett. }}
\def \mpl{ {\nineit  Mod. Phys. Lett. }}
\def \prl{ {\nineit  Phys. Rev. Lett. }}
\def \pr { {\nineit  Phys. Rev. }}
\def \cqg{ {\nineit  Class. Quant. Grav. }}
 \def \ij{ {\nineit  Int. J. Mod. Phys. }}

%%%%%%%%%%%%%%%%%%%%%%%%%%%%%%%%%%%%%%%%%%%%%%%%
\newsec{ Introduction }
%%%%%%%%%%%%%%%%%%%%%%%%%%%%%%%%%%%%%%%%

One of the few universal predictions of string theory is the existence of a
scalar field  (dilaton)  which is  coupled  to matter. The presence of
the dilaton  along with the graviton (and the antisymmetric tensor) in the
string theory effective action was first  pointed out in ref.1. The analogy
with $\omega =-1$ Jordan-Brans-Dicke  theory
$$ S =  \ha \int d^D x \sqG \      [   \eta R   - \omega \eta^{-1}
 (\pa \eta)^2  ] + S_m[G,\psi] $$ $$  =
\ha \int d^D x \sqG \         e^{-2\p}[ R \  - 4\omega
 (\pa \p)^2 ] + S_m[G,\psi] \ \ , \eq{1.1} $$
and an apparent conflict with
 solar system observations (imposing the constraint $\vert \omega \vert >
500$, i.e.  rulling out the presence of a scalar component of gravity) was
noted.

 The full non-polynomial structure of the dilaton couplings in the tree
level string effective action   (which was difficult to determine using
the methods of ref.1)  was first inferred  indirectly  after the  construction
of  $D=10$ supergravity$^2$  (interpreted as a zero slope limit of a
superstring
theory$^3$) and later understood from string theory$^{4,5}$. In contrast to the
original JBD action, the string action contains the dilaton couplings to  other
``matter" fields  already in the `Jordan' or `string' frame
$$ S =  \ha \int d^D x \sqG \  \e{-2\p}   \ [  \ R \ + 4 (\pa \p )^2
    - {1 \over 12} H^2_{\l  \mu \nu}-{1 \over 4} F^2_{\mu
\nu }$$ $$  - { 2 V}(\p)
- (\pa \psi)^2 - m^2 \psi^2 +  ...\ ] \ \ .\eq{1.2} $$
For  generality we have included a dilaton potential term
and also a scalar field  with a  tree level mass $m$.
The corresponding action in the Einstein frame  is
$$ S_E =    \ha \int d^D x {\sqrt { -g }}   \ [  \ R  -2p (\pa \p )^2
     - {1 \over 12}\e{-4p \p }
  H^2_{\l  \mu \nu}  -{1\over 4} \e{ -2p \p } F^2_{\mu \nu} $$ $$   -
{2 \hat V}(\p)
- (\pa \psi)^2 - \e{2p\p }  m^2 \psi^2 + ...\ ] \  \ ,\eq{1.3} $$
$$  g_{\mu \nu }= \e{-2 p\p } G_{\mu \nu }\ \ , \ \ \ \
{ \hat V}(\p) = { \e{2p\p } }  {  V}(\p) \ \ ,
\ \ \ \ p\equiv 2 /( D-2) \ .
\ \eq{1.4} $$
The Einstein frame form of the JBD action (1.1) is
$$ S_E' =    \ha \int d^D x {\sqrt { -g }}   \ [  \ R  -  q_0 (\pa \p )^2
     -{1\over 4} \e{   q_1\p } F^2_{\mu \nu}  -
 e^{q_2\p }(\pa \psi)^2 - \e{q_3\p }  m^2 \psi^2 + ... \ ] \  \ ,\eq{1.5} $$
$$ q_0 = 4\omega + 2p(D-1)\ , \ \ \ q_1= (D-4)p\ , \ \ \  q_2= 2 \ , \ \ \
q_3= Dp \ . \eq{1.6} $$
The actions (1.3) and (1.5) belong to a general class of actions describing
interactions of a massless scalar ``universally" coupled to matter.  While in
(1.3) the dilaton  coupling constant is of order one (there is no
``hierarchy" of scales of  the dilaton and gravitational couplings   at the
string tree level)  it is the value of $\omega^{-1/2}$ that  sets the scale  of
dilaton couplings in (1.5).

There are  two basic types of observational restrictions on  a scale
 (denoted by $\omega^{-1/2}$)  of interactions of a
 massless scalar  with matter (see Sec.2.3 and references
there). The first comes from solar system post-newtonian experiments like radar
time delay measurements.  The second  is related to a cosmological evolution of
the scalar field. If the scalar changes with time, this  produces  a time
variation  of effective particle masses, or, equivalently, in the Jordan frame,
of the gravitational constant. The latter variation can be constrained  from
consideration of
 primordial nucleosynthesis. To satisfy  both post-newtonian
and nucleosynthesis bounds   $\vert \omega \vert $ in (1.1) should be greater
than a few hundreds.  Much stronger constraint applies in the case when the
scalar (like the  string theory dilaton or  one  combination of it  with
moduli)
is coupled  to the  gauge field  kinetic terms in the action. The time
variation
of the electromagnetic coupling should be extremely small to satisfy bounds on
stability of nuclear isotopes.  In the simplest model this gives the
restriction
$\vert \omega \vert > 10^{7}$.

The  crucial  point that may  help to avoid  conflict with observations  is
that
non-perturbative string corrections$^{6,7}$  should modify the structure of low
energy interactions of the dilaton (and of other massless scalars  which  are
present in a $D=4$ string spectrum).
In particular,  a non-trivial dilaton potential (and
hence a mass for  a fluctuation near a minimum)  should be generated. Also,
the
matter mass terms should have a non-perturbative origin    so that their
coupling  to the dilaton
($\sim
exp(-a e^{-2\p})$) should be different  from that  in  (1.3).

 While the dilaton is most probably ``frozen"  at the minimum of its potential
at the present time, it could have played an important  role in  early
Universe.
It seems  reasonable  to  assume that the  potential term becomes  essential
(and supersymmetry is broken) only at rather  late  stage of evolution so that
at
earlier times the dilaton  can be treated as  massless (and is described
approximately by  the tree level effective lagrangian).  Then one  can try to
analyse  possible models of string cosmology  using a ``phenomenological"
approach, i.e.  accounting  for the   effects of a gas of string modes
by adding some ``matter" terms in the tree level metric - dilaton action. We
shall discuss a cosmological scenario$^{8,9}$ based on such
 an approach  and  the issue of its  correspondence with the `standard'
cosmology
in Sec.2.

One may  also hope to gain some useful information about string cosmology by
studying {\it time dependent} solutions of the vacuum metric - dilaton
equations  and  their exact conformal field theory generalizations. While
looking for ``cosmological" conformal field theories  may be of limited
importance\footnote {$^a$} { A conformal theory
corresponds only to a  perturbative (classical) solution of a  superstring
(Bose
string) theory. It is not known whether non-perturbative  solutions (e.g.
extremals of an effective action which contains non-perturbative corrections
like a  dilaton potential) can be described in terms of 2d conformal theories.}
this approach   may be ``complementary" to the ``phenomenological" one (based
on
a low energy effective action containing only leading terms in the $\a'$
expansion but including non-perturbative corrections and ``matter" terms).  We
shall  review some recent results about time dependent  solutions of the string
tree level equations$^{10}$ in Sec.3.

\newsec { Cosmological Scenario}

The ``standard" cosmological scenario  based on  an inflationary phase followed
by a hot Universe phase$^{11}$ does not give answers some basic ``initial
condition" questions like why the Universe was {\it expanding} and  why this
expansion was taking place in {\it three} spatial dimensions.  One may expect
that  the string theory  being a fundamental theory   should
provide answers to such questions. If the basic objects are closed strings it
is  natural to define them in a compact space where they can wind around
possible non-trivial cycles and thus have additional solitonic ``winding"
states
in their spectrum. Flat space is then considered as a limiting case of a torus.
Our  starting point for a discussion of string  cosmology  will be to assume
that
all spatial dimensions are compact and are of characteristic string (i.e.
Planck)
scale ${\sqrt  {\alpha'}}  \sim M_P^{-1}$. The  aim is then to understand why
only three dimensions have expanded while others  remained ``internal", i.e. of
Planck size. A particular  mechanism$^8$  which may provide an explanation of
 why only three (or less) dimensions
are likely to expand  is based on the fact that the  winding
 modes  oppose the expansion.$^{8,9}$

If the expansion did happen the Universe should eventually reach a radiation
dominated phase in which all massive string modes have decoupled. Later on,
when
non-perturbative corrections to the string effective action will become
important  and supersymmetry will be  broken  the Universe we may  enter an
inflationary phase (with one of the scalars in the low energy spectrum playing
the role of an ``inflaton").

\subsec {Basic assumptions  }
Our basic assumptions will be the following:

(i) {\it weak coupling}:  string interactions are small; the dilaton, i.e the
effective string coupling should not increase with time.

(ii) {\it adiabaticity }:
the metric and dilaton are evolving slowly with time so
that higher derivative terms in the effective action can be ignored.

(iii) {\it space is a torus}:  spatial dimensions are periodic so
that winding modes are present in the string spectrum.

\noindent
We shall assume also that the metric and the dilaton depend only on time
$$ ds^2 = - dt^2 + \sm a_i^2 (t) dx_i^2  \  \  , \eq{2.1} $$
$$ a_i= {e}^{\l_i(t)} \ , \ \ \p=\p(t) \ , \ \ N=D-1\ \ . $$
It is useful to introduce the ``shifted''  dilaton
field  $\vp$ (which is invariant under the duality
transformations$^{12,13,14}$)
$$ \vp \equiv 2\p - \sm \l_i \ \ ,\  \ \ \sqG\   \e{-2\p} = \e{-\vp} \   \ .
\eq{2.2} $$
Then the action for the gravitational degrees of freedom interacting with
string
``matter"
$$ S = \ha \int d^D x \sqG \  \e{-2\p}   \ [ \ c + R  + 4 (\pa \p )^2
\ ]  + S_m [ G, \p ]\ , \ \eq{2.3} $$
takes the form
 $$S =  \ha \int dt \ \e{-\vp} [\  c + \sm \dl_i^2 - \dvp^2  - 2U (\l_i ,
\vp)]\  \ \ .   $$
For generality we have included the central charge  deficit term  which is
 set equal to zero in the most of the  present section.
The resulting equations for the scale factors and the dilaton
are$^{13,14,9,15}$  $$  c - \sm \dl_{i}^2  + \dvp^2   = 2 \CE  \ \ , \ \eq{2.4}
$$ $$ \ddl_i - \dvp \dl_i  =   - {\pa \CE \over \pa \l_i }   \ \ , \eq{2.5}
$$ $$ \ddvp - \sm \dl_{i}^2 =    {\pa \CE \over \pa \vp }   \ \ . \eq{2.6}
$$
Eq. (2.4) is a constraint which is conserved  as a consequence of (2.5),(2.6).
This is the generic form  of cosmological  equations after all other possible
variables (like  matter fields, temperature, etc) are elimin.ated. Consider,
for example,  the case when the matter part of (2.3) is represented by a
 (one-loop) free
energy of a gas of string modes in thermal equilibrium at temperature
$\b^{-1}$,
 $$\ \ \  S_m = -\int dt   \ F ( \l , \b  )
\ .$$ Then
$$  c - \sm \dl_{i}^2  + \dvp^2   =  2 \e{\vp} E  \ \ , \ \eq{2.7}  $$
$$ \ddl_i - \dvp \dl_i  =    \e{\vp }  P_i \ \ , \eq{2.8}  $$
$$ \ddvp - \sm \dl_{i}^2 =   \e{\vp } E   \ \ , \eq{2.9}  $$
where
$$  E= F + \b {\pa F \over \pa \b} \ ,  \ \ P_i
= - {\pa F \over \pa \l_i } \ , \
 \ \dot E + \sm \dl_i P_i = 0 \ . \eq{2.10} $$
  Since
$F=F(\l (t) , \b (t) ) \  $  eq.(2.10) is equivalent
to the conservation of the entropy ${ s} = \b^2 \pa F/ {\pa \b} $, $\ \
\dot E + \sm \dl_i P_i = {\dot { s }} / \b $ .
Solving the adiabaticity condition one can in principle express the
temperature in terms of $\l_i$ . Then eqs. (2.7)--(2.9)  take the
form of (2.4)--(2.6) with $$\CE= { e}^{\vp}E (\l , \b (\l) ).$$

The system (2.4)--(2.6) has an obvious mechanical interpretation: it
describes a particle  moving in the potential $\CE$. Since  the dilaton should
not grow during the expansion of the spatial dimensions
$\dvp$ should be negative  ($\dvp$ does not change sign if the r.h.s. of
(2.6) is positive). Then the dilaton term in (2.5) can be interpreted as a
friction force.
%%%%%%%%%%%%%%%%%%%%%%%%%%%%%%%

\subsec { String  phase   }
Let us now assume that a toroidal universe is filled with  classical
  strings in  momentum and winding states which are out of thermodynamical
equilibrium. Since the masses of momentum modes ($\sim a_i^{-1}$) grow  with a
decrease of $\l_i$ while the masses of winding modes ($\sim a_i$)  grow with an
increase of $\l_i$ the ``energy" $\CE$ in (2.4) will grow at both positive and
negative $\l_i$ (and will not depend on $\vp$ since we are discussing classical
contributions). This is easy to see, for example, representing momentum and
winding states  as  ``tachyonic" scalar string modes$^{15}$
$$ S_m = \ha \int dt\  \e{\vp} \  [ \  \dpss -
(\sm \mi \e{-2\l_i} -4 ) \pss
   + \dtpss - ( \sm \tmi \e{2\l_i} -4 ) \tpss  + ...\ ] \ . \
\eq{2.11} $$
Solving the system (2.4)--(2.6) with such $\CE(\l_i)$ one
concludes$^{9}$ that  $\dvp$ remains negative if it was negative at the initial
moment and the trajectory  of the system on the energy -- $\l_i$ plot is going
down (reflecting from the walls of the potential $\CE$) towards the minimum of
$\CE$ at $\l_i =0$. As a result,  $a_i(t)$  are  oscillating near the Planck
scale with the amplitude of oscillations  decreasing   because of the
dilaton damping.

Since the presence of classical winding modes prevents penetration to large
radius region they  must first ``annihilate" (winding and anti-winding state
hitting each other and producing a momentum state) to make  the expansion
possible.  The interaction of classical  strings occurs only when their world
surfaces intersect  but such process is most probable when the number of space
-
time dimensions is less or equal to 2+2=4. This is the idea of the
mechanism$^8$
suggesting an explanation of why only three (or less) spatial dimensions can
expand to ``macroscopic" sizes. If, by  a  fluctuation, some three dimensions
started expanding, winding modes will start annihilating and that will make
further  expansion  possible.

After the expansion had happened,  the universe  enters the second stage of
evolution  in which the ``matter" is represented by a gas of string modes
(defined on an $N$-torus) which are  in thermal equilibrium.
$N$ is  now  the number of expanding spatial dimensions (i.e. three) and for
simplicity we shall assume that all $\l_i$ are equal ($\l_i$ =$\l$).
 The properties of the corresponding ``energy" function $ E(\l) = E (\l,
\b(\l))
$ were studied using the microcanonical ensemble$^{8,16}$ and can be summarised
as follows. Containing the contributions of both momentum and winding modes,
this function (as well as the partition function  and the temperature) is
duality symmetric $E(\l) = E(-\l)$.  It  reaches its maximum in the  `Hagedorn
region' near $\l=0$  where it is almost constant. For large enough $\l$ the
temperature drops below  the Hagedorn temperature $T_H$ and  the massive string
modes go out of thermodynamical equilibrium
($T_H$ is of the same Planck order as the masses of the string modes).
The behaviour of $E(\l)$  at large $ \l$  is  thus   the same
as in the `radiation dominated  region' (where only the massless string modes
contribute to the partition function), i.e.  $E(\l)$ exponentially goes to
zero, $E\sim T^{N+1}a^{N} \sim { e}^{-\l} .$

Though the transition  from the Hagedorn  to the  radiation dominated region is
not fully understood, it is possible to solve the resulting system (2.4)--(2.6)
or (2.7)--(2.9)
$$ c-N\dls +\dvps = 2\e{\vp} E \ \ , \eq{2.12} $$
$$ \ddl -\dvp \dl = -N^{-1} \e{\vp} {\pa  E \over \pa \l } \ \ ,
\eq{2.13} $$ $$ \ddvp - N \dls =  \e{\vp} E  \ \ , \eq{2.14} $$
separately in the two regions and to match the resulting solutions$^9$.
Since in  the Hagedorn era $E=const$, i.e.   the ``potential" is flat and
nothing prevents the expansion to continue. One finds  that the dilaton is
decreasing  while the  scale factor  is  growing  with  a slow-down,
asymptotically approaching a constant
$$ \vp = \vp_0 - {\ln  } \vert  {t^2 -  b^2  } \vert \ \ , \ \
\l=\l_0 + {1\over \sqrt N }    {\ln }  \vert {{t- b }
\over { t+ b }} \vert \ \ . \eq{2.15} $$
Expanding, $\l$ naturally reaches the intermediate region where $E$
drops and the system starts  `rolling down' over the potential  in the
direction
of large  $\l$. In the radiation dominated era
$$P=N^{-1} E\ ,\ \ \  \ E=E_0 {e}^{-\l} \ , \ \ \ \ \rho = E_0
e^{-(N+1)\l}       $$
 there exists a special `power
law' solution with a constant dilaton $\p$ $$  \ \ \l = \l_0\ + \ q\  {\ln}\  t
 \
, \ \ \  \vp = \vp_0\ +\ s\  {\ln }\ t \
 \ , \eq{2.16} $$
$$\ \ q = 2 /(N+1) \ ,  \ \ \ \ s=- 2N /(N+1 )  \ , \ \
 \ \ \ \p= \ha (\vp + N
\l ) = \const  \ \ .  $$
The important point is that this solution is  an  $attractor$ $^9$, i.e. all
solutions (with  the initial conditions $\dvp < 0 $ and $\dl > 0)$
 approach it asymptotically. The conclusion is that without
any  unnatural fine tuning  we  reach
the standard `radiation dominated' cosmological era  with  the dilaton
remaining
 $constant$ at late times.

\subsec {Transition to `standard' cosmology: dilaton
potential and inflation   }

In the radiation dominated era  the massive
 string modes  have already decoupled  and  the dynamics
 is governed essentially by the low energy  effective field theory for
the ``light" fields. Then non-perturbative corrections should become important
and, in  particular, a supersymmetry  breaking phase transition
(generating a potential for  the dilaton) should
happen at some time.

The effect of  a   non-perturbative dilaton potential on
the  cosmological evolution was studied e.g. in refs.17,18,19.\footnote{$^b$}
{  Cosmological solutions  in the presence of an
explicit mass term for the dilaton were  first discussed in ref.20. }
In addition to the dilaton, the effective  action  may include other scalar
modes as well (which may play an important role in a possible inflationary
phase, see below). In general, a $D=4,  N=1$ supersymmetric  effective
action can be parametrised  by the scalar fields $(S,T)$  of  chiral
supermultiplet(s)  and by a Kahler potential and superpotential$^{21,6}$.
In the case of
the  model corresponding to the heterotic string compactified  on a
 Ricci flat  6 -  dimensional space (e.g. torus) with  a  time - dependent
scale $ b =
e^{\s (t)}$
 $$ Re  S = e^{ -2\P}\  , \ \  \ Re  T = e^{  \s/2} \ , \
\ \ \P \equiv \p - 3 \s \ ,\ \ \ \vp\equiv 2\p-3\l-6\s=2\P-3\l \ .
 \eq{2.17} $$
The effective action resulting from compactification on a 6-torus  has the
following form in the string frame
$$ S =  \int d^4 x \sqG \  \e{-2\P}   \{\  \ha  [  \ R \ + \ 4 (\pa \P
)^2  - {6} (\pa \s )^2 \    -  2{ V} (\P, \s)\ ] +  L_m \ \}  \ , \ \eq{2.18}
$$
where  $L_m$  contains  the contributions of the ``axions" (imaginary parts of
the scalars $S$ and $T$) as well as other ``light" matter fields. It is $\P$
that we shall
plays the role of  the dilaton  in the $D=4$ theory.  For the isotropic
spatially flat 4
- metric  $ ds^2 = - dt^2 + e^{2\l(t)}dx_idx^i$
  we get ($N=3$)
 $$S = \ha \int dt \ \e{-\vp} [ N\dl^2 - \dvp^2  + {6} {\dot \s}^2
 - {2 V} (\P, \s) \  + ... \ ]\   \eq{2.19}
$$
($\s$ plays the role  of 6 of   $\l_i$ in (2.3)--(2.6)).
The resulting system of equations can be found from (2.4)--(2.6).
One should also
include the contributions of the energy and pressure of the radiation (see
(2.12)--(2.14)).  The evolution of $\s$ is not important and can be ignored
in the first approximation.

 The  potential  corresponding to the
supersymmetry breaking due to  gaugino condensation and a non-trivial
antisymmetric tensor (axion) background   has the following dependence on $\P$
(see  refs.7,22  and references therein)
$$ V= \sum_{i} exp(-a_i e^{-2\P}) (A_i + B_ie^{-2\P} + C_ie^{-4\P}) \ .$$
In the case of the  two gaugino condensates$^{22}$
$$ { V} = d_1^2\ e^{-a_1 Y }\ [ (a_1Y
+1)^2 -3]\  + \  d_2^2\ e^{-a_2 Y } \ [ (a_2Y +1)^2 -3] $$ $$ -\ 2d_1 d_2 \
e^{\ha(a_1 + a_2) Y }\ [a_1a_2 Y^2 + (a_1 + a_2) Y -2 ] \ , \ \ \ \ \ \ \
Y\equiv e^{- 2\P} \ . \eq{2.20} $$
The constants $d_i$ and $a_i$ depend on
a  gauge group  of the hidden sector (for example, $d_i\sim 10^{-2}$,
 $a_i\sim 10$).   This
potential  starts from zero in the  weak coupling region of large  negative
$\P$,  grows and reaches a local maximum, then decreases to a local minimum
(with negative $V$), then has the second  local maximum and finally goes to
$-\infty$ at  large positive $\P$. Since the potential has a local minimum it
may fix the value of the dilaton. This would   suppress  time variations  of
effective masses and couplings and give  a mass  to the fluctuating part
of the dilaton.

To study the approach to the constant regime and  the  correspondence with the
`standard' cosmology it  is natural to use the Einstein frame.
 A  cosmological  background  in the \sf theory (1.2)
$$ds^2 = \Gmn dx^{\mu }
dx^{\nu }= - dt^2 +  \e{2\l (t) } \ d\Omega^2 \ \ ,  \ \ \P = \P (t) \ \ ,
\eq{2.21} $$
  corresponds to the following background   in the \ef theory (1.3)
$$ds^2_{\E} = g_{\mu \nu } dx^{\mu } dx^{\nu }= e^{-2p\P (t) }
(- dt^2 +  \e{2\l (t) } \ d\Omega^2 \ ) \ \ ,  \  \eq{2.22} $$
$$ ds^2_{\E} = - d\t^2 +  \e{2\L (\t) } \ d\Omega^2   \
\ \   , \eq{2.23} $$
$$d\t = dt \ e^{- p \P (t) } \ , \ \ \  \ \L \equiv \l - p \P  \ ,
\ \ \  p = 2/ ( N-1) \ .  \eq{2.24} $$
 $d\Omega^2$ is the interval of a maximally
symmetric 3 - space with curvature $k$. The cosmological equations in the
Einstein frame  have the form ($N=3$; cf.(2.4)--(2.6),(2.12)--(2.14))
$$ N(N-1) { \dot \le }^2 -  4 (N-1)^{-1}{\dot \P}^2 = 2 \CU  \ \ , \eq{2.25} $$
$$ (N-1) { \ddot \le } + 4( N-1)^{-1} {\dot \P}^2
= {  N}^{-1} {\pa \CU \over \pa \le} \ \ , \eq{2.26} $$
$$ {\ddot \P} + N { \dot \le }{\dot \P} =- {1\over 4} (N-1) {\pa \CU \over \pa
\P }  \ \ ,  \eq{2.27} $$
$$ \CU (\le , \P )  \equiv  \e{4\P /(N-1) } \ [ \ U (\l , \P ) -  c \
]  \ \ .  \eq{2.28} $$
Here  the dots  denote derivatives over the Einstein frame  time $\t$.
Note that the structure of eqs.(2.26),(2.27) is different from that of the
string frame equations (2.5),(2.6). The dilaton $\P$ is not damping the
evolution of the scale factor $e^\L$; it is  expanding  $\L$ that  provides a
friction term in the equation for the dilaton (2.27).

 In general,  $U$ in (2.4) may  contain the contributions of the
spatial curvature, antisymmetric tensor  background ($H_{ijk} = h
\epsilon_{ijk} $ in the string frame), radiation ($\rho = E_0 e^{-(N+1)\l }$)
and the dilaton potential,
 $$ U= - \ha kN(N-1) \ e^{-2\l} \ + \ \fourth h^2
\ e^{-2N\l} \ +\  E_0 \ e^{2\P-(N+1)\l}\  + \ V(\P)  \  \eq{2.29} $$
 (one  can  also include
contributions due to scalar non-relativistic matter$^{23,18}$).
Then
$\hat U $ in (2.28)  is given by
 $$ {\hat U}= - \ha kN(N-1) \ e^{-2\L}\  +\
\fourth h^2 \ e^{-4\P -2N\L} \ +\  E_0\  e^{-(N+1)\L} \  $$
$$ + \ \e{{4\P /( N-1) }} \ { V} (\P) \  - \ c \ \e{{4\P /( N-1) }} \ \
. \ \ \eq{2.30} $$
  In the absence
of a dilaton potential and  radiation    the  asymptotic solution of
(2.25)--(2.28) is the same as in the $\hat U=0 $ case, i.e. a `power law'
expansion and dilaton changing logarithmically in time, $$ \L = \L_0 + N^{-1}
 \ln  \t \ , \ \ \ \P = \P_0 +  \ha N^{-1/2}(N - 1)  \ln  \t  \ \ .
$$ Such  a  dilaton
behaviour would  produce   unacceptable variations of particle masses
and couplings. As we have  discussed above, if the universe is dominated by the
radiation the dilaton approaches a constant value$^{9}$ (see also refs.23,18).
However, the radiation dominated era cannot last  forever. As the universe
will enter the matter dominated era  the dilaton will eventually restart
changing  with time  if it is not suppressed by a potential$^{18}$. The
conclusion is  that to  avoid conflict with observations (in particular, with
the nucleosynthesis bound)
the dilaton should be already  ``fixed" by a
potential  at the time  the universe enters the matter dominated phase.

When   the  non-perturbative  dilaton
potential is ``turned on" during the radiation dominated era   $\P(\t)$
goes through a transitional period  and starts approaching  a (different)
constant corresponding to the  minimum of the potential. As it is clear from
(2.25), if $V$ and hence $\hat U$ (i.e. the effective cosmological constant)
is
positive at the minimum,  the solution with $\P$ sitting at the minimum and the
scale factor exponentially expanding is a stable one (the expansion of the
universe  rapidly damps  the  dilaton oscillations near the minimum).

 The   negative value of  the potential (2.20)  at the
minimum suggests that  the minimum may be unstable$^{19}$ (with the universe
eventually starting contracting and the dilaton starting moving away from the
minimum). However,  the contributions of other scalars  (matter
fields) may  shift the value of the effective cosmological constant making
it  zero or positive.  In the
first case the dilaton relaxes to the minimum while the  universe expands
according to the radiation era law ($e^\L \sim \t^{2/(N+1)}$).  The exponential
inflation one finds in the second case  can  be also achieved of course
by   artificially fine tuning the value of $c$ in (2.30)  to make  ${\hat
U}_{min} > 0$ $^{24,18}$.

%%%%%%%%%%%%%%%%%%%%%%%%%%%%%%%%%%%%
A possibility to have  an inflationary period strongly depends on  details of
supersymmetry breaking  and  structure of non-perturbative  terms  in  the
string low energy effective action (see refs.25,26 and
references therein).\footnote {$^c$}
{For some other papers on inflation in string
cosmology see  refs.27,28.}
Among the conditions necessary  in order to have  a period of
inflation is the existence of a  local minimum
in the potential of some scalar matter field (``inflaton"). Let us
assume   that $\s$ in (2.18)  corresponds to a
flat direction of $V$  and  that the potential $V(\P)$ has a minimum, i.e.  it
generates  a mass term for the dilaton,
$\ V \sim  (\P -\P_*)^2 +... \ $ (we shall assume that $V_{min}=0$).
Considering the model of ref.25 in which  $L_m$ contains a minimally coupled
scalar field $\psi$ with a Higgs-type potential    we get the following action
 in the Einstein frame ($N=3,\  p= 1$ ; the gravitational constant is set equal
to one)
 $$ S_E =  \int d^4 x {\sqrt -g} \     \{  \ \ha R \ - p (\pa \P )^2
-  3 (\pa \s )^2 \   $$
$$   - \ e^{2p\P}\  [ \ \m^2 (\P -\P_*)^2  + \ha (\pa \psi)^2  +  ( M^4 -
\ha m^2 \psi^2 + \fourth h \psi^4 ) + ...  \ ] \  \} . \ \eq{2.31} $$
Here  $\mu$ is related to a supersymmetry breaking scale
and $M$ is of order of a GUT scale  (a possible choice of
parameters is $\mu \sim 10^{-6}M_P$, $\ M \sim 10^{-3}M_P , \ m\sim 10^{-4}M_P
$).
 If initially $\P\not=\P_*$ and the  inflaton $\psi$ is in  the metastable
minimum of its potential ($\psi =0$) the effective cosmological constant
contains two  relevant contributions:$\  O(\m^2)$ and $O(M^4)$. The first term
leads to  a period of chaotic inflation  which is followed (after the dilaton
relaxes to its minimum) by a period  of ``old" inflation driven by the second
term$^{25}$ (for a similar two - scalar inflationary model see ref.29).  The
dynamics of $\s$ and $\psi$ is mostly irrelevant.  Solving the equation for
$\s$  we get an extra $ O(e^{-2N\L})$ term in $\hat U$ in (2.30). Ignoring all
$O(e^{- n\L})$  terms $\  (n > 0) $ in $\hat U$ we find
 $$ {\hat U}=    e^{{ 2p\P }} \  [ \ \m^2 (\P -\P_*)^2  +
M^4   \ ]  +... \ \ . \eq{2.32} $$
The above conclusion about the inflationary periods then follows from the
analysis of  solutions of the system (2.25)--(2.27).

 Note that $M^4$ plays the role of a  $negative$ central charge deficit $c$
in (2.28).  If the dilaton potential term in (2.32) (which  eventually
fixes the dilaton) was not included,  we would  have  found that
asymptotically  the  solutions approach the well-known  ``linear dilaton, flat
string frame metric"
 solution$^{30}$ which in the Einstein frame  takes the
form$^{31}$
 $$ \L = \L_0 \ + \
\ln   \t \ , \ \ \ \P = \P_0  - \ha (N-1)   \ln  \t  \ \ , \eq{2.33} $$
$$ (N-1)^2 = - c \  e^{{4\P_0  / (N-1) }}= M^4 \  e^{{2p \P_0 }} \ .
$$ With $O(e^{- n\L})$ terms in $\hat U$ included,  the solution (2.33) is  an
attractor  only at infinity, i.e. the scale factor will be growing  rather
slowly
(${\dot a } < 1 $)  so that  the horizon problem will not be solved$^{32}$. The
dilaton potential is thus  necessary  also in order to  get  a sufficient
inflation in this model.

A different  mechanism  for  a  realisation of a period of (extended$^{33}$)
inflation  was considered in ref.26 (cf. ref. 28).
Here the main role  is played by a scalar $\s$ corresponding to
a flat direction of the dilaton - moduli potential. The existence of a flat
direction  is a necessary condition  for extended inflation. Though
in general  the  scalar corresponding to a flat direction  may be a non-trivial
combination of $\P$ and the modulus  ($\s$  in (2.17)) we shall use the same
notation ($\s$ for a flat direction and  $\P$ for the ``orthogonal" one)  as in
(2.31).  The new element as compared to the model (2.31) is that we shall
include possible couplings of $\s$ to matter. Let us assume for simplicity that
$\s$ couples exponentially to the kinetic terms and masses of the matter
fields$^{6,22,26}$.
Then the relevant part of the  effective action  is given by (we shall ignore
the
dependence on $\P$ since  it is  fixed by the potential) $$ S_E =
\int d^4 x {\sqrt -g} \     \{  \ \ha  R \ - {3} (\pa \s )^2 \
 - \ \ha \sum_{n} [\  e^{\g_n \s} (\pa \psi_n)^2   +
  e^{\b_n \s} m^2_n \psi^2_n\  ] + ... \} . \ \eq{2.34} $$
In  a   matter dominated  phase the  kinetic terms are not
important while the mass terms produce  a potential for $\s$ given by a sum of
exponentials. If we  further assume that  the potential is dominated by one of
the terms   $e^{\b \s}  \ (\b=\b_i) , $ the resulting system
is similar to  (2.30),(2.31),(2.32)  with $
 \s$  playing the role of $\P$  and the matter density -- the role of
$-c$ (with  this identification $\b={\sqrt  6} $ in
(2.30),(2.31)).\footnote {$^d$}
{ In general, one can  trade one of the  coefficients
in the exponentials  for the JBD constant   in (1.1) by making a Weyl
transformation, i.e. by  going into the Jordan frame  as in  ref.26. }
   The  cosmological equations have  again the  `power law'
solution
 $$ \L = \L_0 \  +  12 \b^{-2}   \ln  \t \ , \ \ \ \s = \s_0
  - 2 \b^{-1}   \ln  \t  \ \ . \
\eq{2.35} $$
 Since
$\s$ is a massless  scalar field  the  values of the coefficients $\g_n,\ \b_n$
of its couplings to matter are in principle  strongly constrained by the
post - newtonian experiments of radar echo delay$^{34}$.  An additional
constraint comes from the condition  that  there should be no significant time
variation of $\s$, i.e.  of masses,   for a consistency of the primordial
nucleosynthesis scenario$^{26,35}$.
The post - newtonian  bound  need  not  apply   to  the
 coefficients  of those fields  $\psi_n$  in (3.34) which may correspond to a
dark matter$^{36,35,37}$.   This  suggest to identify $\b$ with the
constant  of the  dark matter coupling to $\s$. If it
is the dark matter that governs the cosmological evolution in the matter
dominated era   $\b$ is  still constrained  by the primordial
nucleosynthesis bound$^{36,37,35}$.  The value of $\b$   is also subject to the
condition of getting sufficient inflation$^{37,26}$.\footnote {$^e$}
{ We are assuming that the scalar field
corresponding to the flat direction  is not coupled to the gauge field
kinetic terms in the action (it is  only  one  `dilaton' combination of the
original dilaton and the moduli  that  couples to the  gauge terms in an
essential way). If  this scalar  was coupled to the gauge terms, its time
dependence  would be severely constrained$^{38,26,35}$  by the bound on a
variation of the  electromagnetic coupling$^{39}$.}
 It is not  clear whether  these
constraints  can be naturally satisfied in string models.

\newsec { Time  Dependent   Solutions of String  Tree Level
Equations   }

In this section we shall discuss time--dependent (``cosmological") solutions of
the string tree level equations corresponding to the action (1.2) (with $V= -
 c/2$) or (2.3),
$$ R_{\mu \nu} + 2 D_{\mu} D_{\nu} \p  +
\ha {\a'} R_{\mu \a \b \g} R_{\nu}^{\a \b \g}  + ... = 0 \ \ , \
\eq{3.1} $$
 $$ c +2 D^2 \p - 4 (\pa \p)^2  -
 {1\over 4}{\a'}
R_{\mu \a \b \g}R^{\mu \a \b \g} + ... =0  \ \ . \eqno (3.2) $$
( $c= -{2\over 3 \a' }(D-26)\ $in the bosonic
string).  One of the motivations for
studying time dependent solutions of (3.1),(3.2) is the following.  In the
previous section we were  assuming  the ``adiabaticity" of evolution, i.e.
that
the fields change  slowly  enough so that  higher derivative terms in the
effective action can be ignored. It does not look sensible to  include only a
finite number of terms in the $\a'$ - expansion (once one of them
 gives significant contribution  others should be
 important as well).
To go beyond the ``adiabaticity" assumption  (what may be
necessary  in order to  clarify the issue of cosmological singularity, etc)
 one should really  look for exact (all orders in $\a'$) solutions of the
string
equations. At the level of (super)string perturbation theory this is equivalent
to  finding the   corresponding conformal theories  which admit a
``cosmological" interpretation.

Unfortunately, very  few  examples of such
theories are known at present.  Apart from the trivial (flat metric, linear
dilaton) solution$^{30}$  and the $R\times S^3$ solution$^{31}$ (based  on
$SU(2)$  WZW  theory)\footnote {$^f$}
{  An exact ``time--dependent" solution is represented
of course by any WZW theory$^{40}$ with a group G which has  one non-compact
generator. An   example  is provided by SU(1,1) WZW model$^{41}$
which can be interpreted as the D=3 anti de Sitter space-time with a vector
field background$^{42 }$. It is likely, however, that one  should necessarily
gauge a subgroup  of G (i.e. to  consider a non-compact coset model) to get
rid of the negative norm states$^{43,44}$.} other known exact  solutions
which
are based on  gauged WZW theories$^{45}$ do not look very
appealing as cosmological backgrounds: they have very few
  (abelian) symmetries and are often  singular$^{46-53,9 }$.\footnote {$^g$} {
 One can give a ``cosmological" interpretation
to a static ``black hole" - type  solution by rotating a space-like direction
into a  time-like$^{13,9,50,52,53}$.}
  It  appears  as if they correspond to  a
 rather  small and  special  subclass of $D>2$  solutions of eqs.(3.1),(3.2).
For example, only a subset of the
simplest ``toroidal" cosmological solutions of (3.1),(3.2) with $N=D-1$
commuting
isometries$^{54}$ has  an identified  coset  conformal field theory
counterpart$^{50}$.

 Standard
cosmological backgrounds   have their spatial sections  represented by
{\it maximally symmetric} $N$ - dimensional manifolds (e.g.  a sphere, a flat
space or a pseudosphere). If there are such regular solutions of the
leading order  string  effective equations (3.1),(3.2)  then there should
exist the corresponding ``maximally symmetric"
conformal field theories.\footnote {$^h$} {   The only example of a  solution
with a maximally
symmetric space which has  known   conformal field theory interpretation is
the  ``static" N=3 solution  of ref.31.   N=3 (pseudo)sphere is
special being equivalent to a group space.  Higher dimensional spheres and   de
Sitter spaces do not directly correspond to conformal theories. For example,
the
``(anti) de Sitter string" of ref.44 based on  gauged  SO(D,1)/SO(D-1,1)    (
SO(D-1,2)/SO(D-1,1) )  WZW model with D  larger (smaller) than  26
 has a space-time interpretation not in terms of the (anti) de Sitter
space-time  but in terms of a  background  which does not have a maximally
symmetric
 subspace (see refs.48,49,53).}
The  first step towards  understanding of  some features of  these
hypothetic conformal theories is to  study  general  solutions of the
leading order equations (3.1),(3.2)  which have a high degree of symmetry.
In particular, it seems important to generalise the solution of ref.54 (its
isotropic  limit)  to the case when the spatial sections have a
non-zero curvature$^{10}$.

In what follows we shall  describe  some  known  ``cosmological" solutions  of
(3.1), (3.2)  starting with the most symmetric ones and proceeding in the
direction of  decreasing  symmetry.

\subsec {  Solutions with maximally symmetric space  }

 Let us first note that
the only  $perturbative$ solution of eqs. (3.1),(3.2) with a maximal
$space-time$
symmetry of the metric is the  flat  solution of ref.{30}
$$ G_{\mu \nu } = \delta_{\mu \nu} , \ \  \ \  \p = \p_0 - b t  \ ,
\ \ 4 b^2 = - c  \ . \eq{3.3} $$
In fact, if the space-time curvature is constant the only solution of the
leading order  form of eq.(3.1) is (3.3).  Assuming $\p=const$ and $R\not=0$
one may  hope  to solve eq.(3.1)  by   trying to  compensate  the  leading
order term $R$ by the  $\a'R^2$ - correction$^{55}$.  The two terms cancel each
other
if the curvature is negative, $\a'R=-N(N-1)$.
However, the resulting $\a'R$ is not small so that all higher loop corrections
to
the $\b$-function (3.1) are equally important and   we  fail to find
 a consistent solution\footnote {$^i$} {  In
general, one should not expect a solution to exist since  it is believed that
sigma models with maximally symmetric target spaces have mass generation, i.e.
are not conformal theories.  Still, in the case of a negative curvature
there is a   formal possibility that   the beta function may have a
 non-perturbative zero. }.

Next, let us  consider the cosmological backgrounds with
maximally symmetric $space$  $$ds^2 = - dt^2 +  a^2 (t) \ d\Omega^2 \ \ , \ \ \
\  \ \  d\Omega^2 =  g_{bc}dx^b dx^c\  \  , \eq{3.4} $$  $$ a = { e}^{\l (t)} \
,
\ \ \p=\p(t) \ ,  \ \ \ \ b,c= 1, ..., N \ , \ \ \ \  D=1 + N \ \ . $$
$ g_{bc}$ is a  metric of a maximally
symmetric $N$ - dimensional space  with the radius of curvature $k^{-1} \ \
(k=-1,0,1) $, i.e. $R_{bc}= k(N -1) g_{bc}$.  The flat solution (3.3)
corresponds to
$$k=0\ , \ \ \  \ \l=\l_0 = \const \ , \ \ \  \  \p = \p_0 - b t  \ ,
  \  \ \  \  \ \vp \equiv  2\p -N \l  = \vp_0 - 2bt \ . \eq{3.5} $$
The resulting system of
equations is (2.4)--(2.6), i.e.
$$ c-N\dls +\dvps = 2 U \ \ ,  \eq{3.6} $$
$$ \ddl -\dvp \dl = -N^{-1}  {\pa  U \over \pa \l } \ \ ,
\eq{3.7} $$ $$ \ddvp - N \dls =  {\pa  U \over \pa \vp}   \ \ ,
\eq{3.8} $$
where
$$ U = - k N (N-1) \ e^{-2\l} \ \  .
\eq{3.9} $$
We shall first ignore other possible contributions to $U$ which may  come from
the  antisymmetric tensor and gauge field backgrounds (cf.(2.29)). The simplest
solution corresponds to the flat space sections ($k=0$, i.e. $U=0$)  and is the
isotropic case of the solution of ref.54
$$ \vp = \vp_0 - {\ln   \sh }  {2} bt   \ \ ,\ \ \ \  4b^2 = -c \ ,  \eq{3.10}
$$
$$\l = \l_0  +  {1\over \sn}  {\ln  \th  }bt \ \ , \eq{3.11} $$
i.e.
$$ \p =\ha(\vp + N \l)= \p_0 \ - \ \ha   \ln  [ (\sh  bt )^{\sn +1} (\ch  bt
)^{-\sn + 1}\ ] \ . \eq{3.12} $$
Asymptotically at large $t$ it approaches the flat solution (3.5).

 One  may ask how ``close" can string solutions with $k \not=0$ resemble
 the  maximally symmetric $D$ - dimensional  de Sitter space.
Naively, one  could  expect that the role of $c$ in (3.6)  is similar to that
of
the cosmological constant in the corresponding  Einstein equation.  In fact,
rewriting (3.6) in terms of the original dilaton $\p$  we get
$$   N(N-1)\dls +
4\dops -4N \dop \dl = -c - N (N-1)\ k\ \e {-2\l} \  \ .
\eq{3.13} $$ If $\p=\const \ $ (3.13) has the  usual de Sitter ($c < 0 $)
$$  {\l } =\l_0  + \ln \cosh Ht  \ \  \ (k=+1) \ ;
\ \ \  \  {\l } = \l_0 + \ln \sinh Ht \ \   \ (k=-1) \ ; \eq{3.14} $$
$$ \ \ \ \ \l = \l_0' + Ht \ \  \
\ \ (k=0) \ ,  $$  $$   \ \ c=-  N(N-1) H^2\ , \ \ \ \
 \l_0=- \ln H \ \ \     $$
or  anti de Sitter ($c > 0$)
$$ \ {\l } = \l_0 + \ln \sin Ht  \ \ \ \
\ (k=-1) \ , \ \ \ \  \ c = N(N-1) H^2\ \  \eq{3.15} $$
solutions. The point, however, is that while $\p=const $  and  the (anti) de
Sitter metric solve (3.6) and (3.7) they do not satisfy the remaining dilaton
equation (3.8).
 That is why the dilaton should necessarily change with time
producing  a ``deformation" of the de Sitter metric$^{9,10}$.
 In fact, it turns out that it is the time variation of the dilaton
and not that of the scale factor that ``compensates" for the presence of the
``cosmological constant" $-c$  in (3.13) in the  asymptotic region of  large
$t$.

Solutions of (3.6)--(3.8) with positive $k$ or positive  $c$  appear to be
singular$^{10}$ (though in some cases  the   singularity may be a coordinate
one
as in the anti de Sitter case, the dilaton  always  starts growing in a finite
 period of time making such solutions  unphysical). If $c\leq 0$ and the space
has a negative curvature one finds a regular solution with the dilaton always
decreasing with time (we assume that  $\dot \vp < 0$ at $t=0$). If
$\l$ is contracting at the initial moment  it  eventually reflects from the
potential wall and   expands to infinity. The expansion is with slow-down due
to
the damping effect of the dilaton. The large $t$ asymptotics of the
solution$^{10}$
 is $different$ from (3.5)
$$ \l \simeq \l_1\ + \ \ha   \ln  t \ , \eq{3.16} $$
$$ \  \vp  \simeq \vp_1\ - \ 2bt \  - \ \fourth  N  \ln  t \
,  \ \ \ \ \ \ \p  \simeq \p_1 \ - \ bt \  +  \  \fourth N  \ln   t \
,\eq{3.17} $$
 i.e.  the scale factor is slowly growing
while the dilaton is linearly decreasing as in (3.5) in order to compensate for
the non-vanishing $c$.

Let us now consider the  case of non-vanishing antisymmetric tensor background.
The equation for the   antisymmetric tensor of rank $n-1$   $$
D_{\l_1 } ( \e{-2\p} H^{\l_1 ... \l_n} ) = 0 \ \ . \eq{3.18} $$
 has two classes of   non-trivial solutions consistent
with symmetries of the ansatz (3.4). The first  is found if the number of space
dimensions $N$ is equal to the rank  $n$ of the antisymmetric tensor field
strength$^{56}$

$$ H_{ 0 a_1...a_{N-1} } =0 \ , \ \  H_{  a_1...a_{N} } = h \ep_{  a_1...a_{N}
}
\ , \ \ h=\const \  \ . \eq{3.19} $$
Then $U$ in (3.6) takes the form
$$ U= - \ha  k N(N-1) \e{-2\l} + \fourth h^2 \e{-2 N \l} \ \ .  \eq{3.20} $$
The two particular   cases relevant for string theory correspond to $N=2 $
(vector field background$^{57}$) and $N=3$ (rank 2 antisymmetric tensor
background$^{56}$).

If the spatial curvature is $positive$  ($k>0$)
the potential $U$ has the minimum at $\l=\l_0$
$$ h^2 = 2k (N-1) \e{2(N-1) \l_0} \ \ , \ \ \ \
 U(\l_0) = - \fourth (N-1) h^2 \e{-2 N \l_0} \ < \ 0 \ \ . \eq{3.21} $$
Then if  $c<0$ the system (3.6)--(3.8) has the following  ``static" solution
$$ \l=\l_0 \ , \ \ \ \  \vp=\vp_0 - 2bt  \ ,  \ \ \ \  \p =\p_0 - bt  \ ,
 \ \  \ \ 4b^2 =  |c| - \ha (N-1) h^2 e^{-2 N \l_0}\ . \eq{3.22} $$
For $N=3$  it has the
well known  conformal field theory  generalisation represented by the
direct product of the $D=1$ `time' theory with linear dilaton  and the $SU(2)$
WZW theory (i.e. $S^3$ parallelised by the antisymmetric tensor
background)$^{31}$.  The
$N=2$ case (with $b=0$) was considered, e.g., in ref.57. In this case the
constant $F_{ab}$--flux ``compensates"  for the  curvature of
$S^2$. It is possible to interpret the $SU(1,1)$ WZW model as an exact
conformal
field theory which generalises this solution to all orders in $\a'$
expansion$^{42}$.

 The general solution is regular if  $ c_{ eff} \equiv c -2U(\l_0) \leq 0$.
Then
 $ \dvp $ remains negative if it was
negative at $t=0$, i.e. the dilaton term in (3.7) plays the role of a
damping force. As a result, the solution (3.21) is  an attractor, i.e.  it is
the asymptotic form of  solutions  with $\dvp <0\ $ (the
space-time is asymptotically $R \times S^N$)$^{57,31}$.

If  the spatial curvature  is negative or zero ($k \leq 0$),  $U$
in (3.19) is positive and has no local minima. The  qualitative behaviour of
solutions   is then the same as in the  absence of the antisymmetric tensor
background (3.9), i.e.  for $c \leq 0$ the dilaton is decreasing while the
scale factor expands with slow-down (or first contracts to a minimal  value and
then expands to infinity)$^{10}$.

 The second class of solutions of (3.18)   exists if $N=n-1$ (i.e.
$N=1,2$)$^{56}$
 $$H_{\l_1 ... \l_{N+1}}= h \ \e{2\p } \ep_{\l_1 ... \l_{N+1}} \ \
, \ \ h=\const \ \ . \eq{3.23} $$
Then
$$ U = -\ha  k N(N-1) \e{-2\l} + \fourth h^2 \e{ 4\p} =
  -\ha  k N(N-1) \e{-2\l} + \fourth h^2 \e{ 2\vp + 2N\l} \ \ . \eq{3.24} $$
The antisymmetric tensor contribution to $U$ in this case is equivalent to
 the ``two-loop" term in the dilaton potential.
If the space is flat ($k=0$)  the  system
(3.6)--(3.8)  with  $U$  given by (3.24)  has a simple analytic
solution$^{10}$.  The expression for $\vp$ is the same as
 in (3.10) while $\l$
and  the original dilaton $\p$  are given by
  $$ \l = \l_0 \  - \ { 1 \over N}  {\ln  } [  A^{-1}{({\th
}bt)}^{-\sn} + A \ {({\th }bt)}^{\sn} \ ]\ , \eq{3.25} $$
 $$ \p = \p_0 \ - \ \ha
 \ln  ( \sinh bt\  \cosh bt \
 [\  A^{-1}{({\th }bt)}^{-\sn} + A \ {({\th }bt)}^{\sn}
\ ]\ ) \ , $$ $$ A^2 = {h^2 \over 32
b^2} \e{2\vp_0 + N \l_0 } = {h^2 \over 8 |c|} \e{4\p_0} \ \  $$
(eq.(3.25) reduces to (3.11) in the
limit  of $h=0$).
The large $t$ behaviour of this solution is the same as in (3.5), i.e. the
scale
factor approaches its maximal value ($\l_0 -  N^{-1 }  \ln  (A+A^{-1}))$
while the  dilaton  is linear. This is not surprising  since the
effect of the $O(h^2)$ term in $U$ (3.24) becomes negligible because of the
decrease of the dilaton. In the special case of $c=0$ one finds $$\vp =
\vp_0 \ - \ {\ln \ }t   \ \ , \ \ \ \ \ \ \ \l = \l_0 \  - \ { 1 \over N}
{\ln}  (A^{-1}t^{-\sn}+ A\ t^{\sn}) \ , \eq{3.26} $$  so that the scale factor
 grows at small $t$ until it reaches its maximum at $t_*= A^{- 1/ \sn  } $ and
then asymptotically contracts to zero.The dilaton $\p$ first grows and then
starts decreasing.

When $k < 0$ the asymptotic behaviour of the solution is determined  by the
first
term in the potential (3.24), i.e. it coincides with (3.16),(3.17).
 This conclusion seems
to be valid in the general  case of  $c \leq 0 \ , \ k < 0 $ and a dilaton
potential $V(\p)$ given by a sum of exponentials $\e{ r\p } \ , \ r >0 $
with positive coefficients. In fact, a slow growth of $\l$  and  a rapid
decrease of the dilaton $\p$ with time implies that the dilaton potential term
in $U$ will be negligible  at late times.

We conclude that the are three basic asymptotic regimes of  regular
solutions with maximally symmetric space (dilaton is always linear at large
$t$): (i) $k=0$: flat metric (eq.(3.5)); (ii) $k>0$: a
 sphere of fixed radius ``parallelised" by a background antisymmetric
tensor field strength (eq.(3.22)); (iii) $k <0 $: the non-trivial expanding
($a\sim t^{1/2}$) space-time (eqs.(3.16),(3.17)). It is an interesting question
which  conformal theory  corresponds to the third asymptotics.

Let us note that we have described the solutions in the string frame which is
most appropriate for a discussion of correspondence with conformal theories.
The form  of the solutions in the Einstein frame can be found using
(2.22)--(2.24). It is the rapid (linear)  decrease of the dilaton that
determines the  asymptotic behaviour of the scale factor in the Einstein frame.
As a result,  all  asymptotic solutions (3.5),(3.22) and (3.16) look the same
being transformed  into the Einstein frame. Namely,  if  in the string frame
 $$
\l = \l_0 + q   \ln
 t \ , \ \  \p  = \p_0 - bt \ ,  \  \ \ \vp \simeq \vp_0 - 2bt \  , \ \  \ b >0
\ \ , \eq{3.27} $$
then in the Einstein frame we get$^{31,10}$
$$
\ \ \p = \p_1 - \ha (N-1) \ {\ln   } (\t -\t_0 ) \
, \ \eq{3.28} $$
$$ \L = {\le }_1 +  \ {\ln   } (\t -\t_0 )  + q  \ln  {\ln   }
(\t - \t_0) \ \ . \eq{3.29} $$
While $\p (\t)
$ is decreasing much slower than $\p (t) $ ,  $ \ \le (\t) $ is  still growing
logarithmically with the coefficient of the leading logarithm being
$universal$, i.e. independent of  $b$
in the dilaton $\vp (t) $ or   $q$ in $\l (t) $.  This implies that looking at
asymptotics of solutions in the Einstein frame is not  sufficient in order to
identify different exact solutions corresponding to different conformal
theories.

\subsec { Anisotropic solutions   }

The  simplest class of anisotropic  solutions$^{54}$ is given by the
spatially flat metric (2.1)   (we are
assuming $c < 0$)
$$   \l_i = \l_{i 0} +{
q_i}\  {\ln \tanh \ }bt \ \ , \ \ \ \ \  \sm q_{i }^2 =1 \ \ ,  \eq{3.30} $$
with  the dilaton $\vp$ being the same as in (3.10).
At large times the metric is flat  (scale factors  approach constants) while
the
dilaton is linear (the corresponding  asymptotic solution in the Einstein frame
is again  the isotropic `Milne' universe $ e^\L \sim \t $).
  This
solution can be generalised (e.g. by using duality transformations) to the case
of non-vanishing antisymmetric tensor backgrounds$^{58,50}$.

A subclass of the
backgrounds (3.30) can be identified as representing the leading form of the
exact solutions corresponding to some gauged WZW theories (e.g. $SL(2,R)\times
SO(1,1)^{D-2}/SO(1,1)$ coset models)$^{50,52}$.  Given that the $D=2$
``black hole" metric$^{59,45}$ is related to the  $D=2$  cosmological
solution$^{54}$ by a complex rotation$^{13,9}$ one can  construct, for example,
a $D=4$ cosmological solution by taking the direct  product$^{52}$ of rotated
black hole theory ($SL(2,R)/SO(1,1)$ with negative level number)  with $R^2$.
The
corresponding leading order metric is the following particular case of
(2.1),(3.30) $$ ds^2 = - dt^2 + \tanh^2bt \ dx_1^2 + dx_2^2 + dx_3^3 \ .
\eq{3.31} $$ One can obtain  other anisotropic solutions by taking  various
direct products
of simple  WZW models, e.g. of  the ``rotated" $SL(2,R)/SO(1,1)$
theory  with the $SU(2)/U(1)$ euclidean black hole theory$^{51,52}$.

Less trivial but less symmetric (inhomogeneous) and singular  anisotropic
solutions  correspond to  $SO(D-1,2)/ SO(D-1,1)$ or $SO(D-1,2)/ SO(D-1,1)$  WZW
models$^{44,47,48,49,53}$.  The  metric   and the dilaton of the
simplest $D=3$ model  which solve (3.1),(3.2) in the
leading order approximation  can be represented, for example,  in
the form$^{49}$
 $$ds^2 = - dt^2 + b^{-2}
(x_1^2 + x_2^2 - 1)^{-1} [ \tanh^2 bt\ dx_1^2
+ \coth^2 bt \ dx_2^2]  \ , \eq{3.32} $$
$$ \p  = \p_0
- {\ln \sinh \ }  {2} bt - \ha \ln (x_1^2 + x_2^2 - 1)
  \ \ ,\ \ \ \ b^2=k^{-1}\ , \  \ \  4b^2 = -c \ .  \eq{3.33} $$
One can take a direct product of the $D=3$ model with a gaussian model
to shift the value of the central charge. The large $t$ asymptotics  of
(3.32),(
3.33)
is the product of the time line
and the $D=2$ euclidean ``black hole" background.

\newsec {Concluding remarks}
Though the  dilaton  should be  ``frozen" at the minimum of a
non-perturbative potential at late times ( or large distances) it may play an
essential role in early string cosmology. It is  important to
study the transitional period  during which  the dilaton is  switching from
its perturbation theory regime to a non-perturbative one. To be able to
analyse
in detail this transition and  late time  string  cosmology
  one needs to have   better  understanding  of the non-perturbative  structure
of the low energy string effective action.

 One of the most characteristic
properties of the dilaton coupling is its  damping or stabilisation effect on
cosmological solutions. In general (assuming that the time symmetry is broken
by the initial condition ${\dot \p } < 0$)  the dilaton coupling introduces a
kind of dissipation into the system.
As a consequence, the  second  order string effective equations  reduce to
the first order renormalisation group
 equations in  the case  of large dilaton damping (see e.g.
refs.60,10).   The coupling of the  dilaton
(dark) matter  may lead to an additional source of
matter entropy production$^{35}$.

 As for the tree level exact solutions, it would be
interesting  to  identify  new conformal field theories which may correspond to
the maximally symmetric solutions of the leading order string equations.
The analysis of exact time dependent string  solutions and a test string
propaga
tion
on the corresponding  backgrounds may  shed some light on
 a number of conceptional problems
(like  which metric one should use to measure singularities$^{61,53}$,
expansion$^{31,27,28}$,  etc)
 related to the fact that the
string gravity is described by the metric $and $ a scalar dilaton field.

\vskip .5in
I would like to acknowledge  a financial support of Trinity College,
Cambridge.

\vskip .3in
\vfill \eject

\centerline{\bf References}
\bigskip
\item{1} J. Scherk and J. Schwarz, \np
 B81 (1974) 118; \pl 52B (1974) 347.
\item{2} A. Chamseddine, \pr D24 (1981) 3065;

E. Bergshoeff, M. de Roo, B. de Wit and P. van Nieuwenhuizen,

\np B195 (1982) 97;

G. Chapline and N. Manton, \pl B120 (1983) 105.
\item{3}F. Gliozzi, J. Scherk and D. Olive, \np B122 (1977) 253;

M. Green, J. Schwarz and L. Brink, \np B198 (1982) 474.
\item{4} E.S. Fradkin and A.A. Tseytlin, \pl B158 (1985) 316;

\np B261 (1985) 1.
\item{5} C.G. Callan, D. Friedan, E. Martinec and M.J. Perry, \np
B262 (1985) 593.
\item{6} E. Witten, \pl B155 (1985) 151;

P. Candelas, G. Horowitz, A. Strominger and E. Witten,

 \np B258 (1985) 46.

\item{7} J.P. Derendinger, L.E. Ibanez and H.P. Nilles, \pl B155 (1985) 65;

M. Dine, R. Rohm, N. Seiberg and E. Witten, \pl B156 (1985) 55;

M. Dine and N. Seiberg, \pl B162 (1985) 299.

\item{8}
R. Brandenberger and C. Vafa, \np B316 (1988) 391.
\item{9}
 A.A. Tseytlin and C. Vafa, \np B372 (1992) 443.
\item{10} A.A. Tseytlin, preprint DAMTP-15-1992; \ij D1 (1992) 111.
\item{11} A.D. Linde {\nineit Particle Physics and Inflationary Cosmology}
(Gordon and Breach, New York, 1990);

E.W. Kolb and M.S. Turner, {\nineit The Early Universe}
(Addison - Wesley, New York, 1990).
\item{12} E. Smith and J. Polchinski, \pl B263 (1991) 59.
\item{13} A.A. Tseytlin, \mpl A6 (1991) 1721;

in {\nineit Proc. of the First International A.D. Sakharov
Conference on Physics,

 Moscow  27 - 30 May 1991}, ed.  L.V. Keldysh {\nineit  et al.},

(Nova Science Publ., Commack, N.Y., 1991 ).

\item{14}G. Veneziano, \pl B265 (1991) 287.
\item{15} A.A. Tseytlin, \cqg 9 (1992) 979.
\item{16} N. Deo, S. Jain and C.-I. Tan, \pr D40 (1989) 2626;

\pl B220 (1989) 125; Harvard preprint HUTP-91/A0235, 1991.

M. Bowick and S. Giddings, \np B325 (1989) 631.
\item{17} K. Maeda and P.Y.T. Pang, \pl B180 (1986) 29.
\item{18} N. Kaloper and K.A. Olive, University of Minnesota preprint
UMN-TH-1011, 1991.
\item{19} J. Garcia--Bellido  and M. Quir{\'o}s, preprint CERN-TH.6442, 1992.
\item{20}  G.F. Chapline and G.W. Gibbons, \pl B135 (1984) 43.
\item{21}  E. Cremmer, S. Ferrara, L. Girardello and A. Van Proeyen,

\np B212 (1983) 413.
\item{22}L. Dixon, V. Kaplunovsky and J. Louis, \np B355 (1991) 649;

 J.A. Casas, Z. Lalak, C. Mu{\~n}oz and G.G. Ross,
\np B347 (1990) 243;

B. de Carlos, J.A. Casas and  C. Mu{\~n}oz, preprint CERN-TH.6436, 1992.
\item{23}  S. Kalara, N. Kaloper and K.A. Olive, \np B341 (1990) 252;

J.A. Casas, J. Garcia--Bellido  and M. Quir{\'o}s, \np B361 (1991)
713.
\item{24}S. Kalara and K.A. Olive, \pl B218 (1989) 148.
\item{25}M.C. Bento, O. Bertolami and P.M. Sa, \pl B262 (1991) 11.
\item{26}J. Garcia--Bellido  and M. Quir{\'o}s, \np B368 (1992) 463.
\item{27} N. S\'anchez and G. Veneziano, \np B333 (1990) 253;

M. Gasperini, N. S\'anchez and G. Veneziano, \np B364 (1991)365.

\item{28}  B.A. Campbell, A.D. Linde and K.A. Olive, \np B355 (1991) 146;

 R. Holman, E.W. Kolb, S.L. Vadas and Y. Wang, \pr D43 (1991) 995 .

\item{29} A.D. Linde, \pl B249 (1990) 18.
\item{30}  R. Myers, \pl B199 (1987) 371.
\item{31}I. Antoniadis, C. Bachas, J. Ellis and D. Nanopoulos,
\pl B211 (1988) 393;

 \np B328 (1989) 115.
\item{32} J.E. Lindsey, \cqg 9 (1992) 1239.
\item{33} D. La and P.J. Steinhardt, \prl 62 (1989) 376.
\item{34} C.M. Will, {\nineit  Phys. Rep. } 113 (1984) 345
\item{35}J.A. Casas, J. Garcia--Bellido and M. Quir\'os, \cqg 9 (1992) 1371;
\pl B272 (1992) 94.
\item{36} T. Damour, G.W. Gibbons and C. Gundlach, \prl 65 (1990) 123;

 T. Damour and C. Gundlach, \pr D43 (1991) 3873.
\item{37}R. Holman, E.W. Kolb and Y. Wang, \prl 65 (1990) 17;

G. Piccinelli, F. Lucchin and S. Matarrese, \pl B277 (1992) 58.
\item{38} K. Maeda, \mpl A3 (1988) 243.
\item{39} F.J. Dyson, \prl 19 (1967) 1291;

J.N. Bahcall and M. Schmidt, \prl 19 (1967) 1294.nn

\item{40} E. Witten, {\nineit  Commun. Math. Phys. } 92 (1984) 455.
\item{41} J. Balog, L. O'Raifeartaigh, P. Forgacs and A. Wipf,

\np B325 (1989)
225.
\item{42} I. Antoniadis, C. Bachas and A. Sagnotti, \pl
B235 (1990) 255.
\item{43} L. Dixon, J. Lykken and M. Peskin, \np B325 (1989) 325;

  I. Bars, \np B334 (1990)125.
\item{44} I. Bars and D. Nemeschansky, \np B348 (1991) 89;

I. Bars, in {\nineit Proc. of XX-th Int. Conf. on Diff. Geom. Methods in
Physics},

eds. S. Catto and A. Rocha (World Scientific, 1992);

E.S. Fradkin and V.Ya. Linetsky, \pl B261 (1991) 26.
\item{45} E. Witten, \pr D44 (1991) 314.
\item{46} J.H. Horne and G.T. Horowitz, \np B368 (1992) 444.
\item{47} M. Crescimanno, \mpl A7 (1992) 489.

\item{48}I. Bars and K. Sfetsos, \mpl A7 (1992) 1091;

\pl B277 (1992) 269.
\item {49} E.S. Fradkin and V.Ya. Linetsky, \pl B277 (1992) 73;

 A.H. Chamseddine, \pl B275 (1992) 63.

\item {50}P. Ginsparg and F. Quevedo, Los Alamos preprint LA-UR-92-640, 1992.

\item {51}P. Horava, \pl B278 (1992) 101;

D. Gershon, Tel-Aviv U. preprint TAUP-1937-91, 1992.
\item{52}C. Kounnas and D. L\"ust, preprint CERN--TH.6494, 1992.
\item{53}I. Bars and K. Sfetsos, preprints USC-92/HEP-B1; USC-92/HEP-B2, 1992.
\item{54}M. Mueller, \np B337 (1990) 37.
\item{55} K. Behrndt,  preprint  DESY--92--055, 1992.
\item{56} P.G.O. Freund, \np B209 (1982) 146.
 \item {57} S. Randjbar-Daemi, A.Salam and J. Strathdee, \np B214 (1983) 491;

E. Sezgin and A. Salam, \pl B147 (1984) 47;

K. Maeda and H. Nishino, \pl B158 (1985) 381;

K. Maeda, \cqg 3 (1986) 233;  \cqg 3 (1986) 651.
\item{58} K.A. Meissner and G. Veneziano, \pl B267 (1991) 33;

\mpl A6 (1991) 339;

A. Sen, \pl B271 (1991) 295;

M. Gasperini, J. Maharana and G. Veneziano, \pl B272 (1991) 277;

  \item {59}  S. Elitzur, A. Forge and E. Rabinovici, \np B359 (1991)
581;

G. Mandal, A. Sengupta and S. Wadia, \mpl A6 (1991) 1685.

\item {60}
 S.Das, A.Dhar and S. Wadia, \mpl A5 (1990) 799;

 A. Cooper, L. Susskind and L. Thorlacius, \np B363 (1991) 132.

 \item {61} G.T. Horowitz, in {\nineit General Relativity and Gravitation}
, ed. N. Ashby {\nineit et al.}

 (Cambridge Univ. Press, 1990).

\vfil\eject

\bye